\documentclass[twocolumn,notitlepage,prb,superscriptaddress,showpacs]{revtex4-1}
\usepackage{hyperref}
\usepackage[usenames,dvipsnames]{color}
\usepackage{amsmath}
\usepackage[english]{babel}
\usepackage{amssymb}
\usepackage{graphicx}
\usepackage{epstopdf}
\usepackage{tabularx}
\usepackage{multirow}
\usepackage[]{todonotes}

\begin{document}
\title{Critical behavior of the classical 
spin-1 Ising model: a combined low-temperature series expansion and Metropolis Monte Carlo analysis}


\author{Amir Taheridehkordi}
\affiliation{Department of Physics and Physical Oceanography, Memorial University of Newfoundland, St. John's, Newfoundland \& Labrador A1B 3X7, Canada} 
\author{Roberto Zivieri}
\email{zivieri@fe.infn.it}
\affiliation{Department of Mathematical and Computer Sciences, 
Physical Sciences and Earth Sciences, University of Messina, Messina, Italy} 

\date{\today}

\begin{abstract}
In this paper, we theoretically study the critical properties of the classical spin-1 Ising model using two approaches: 1) the analytical low-temperature series expansion and 2) the numerical Metropolis Monte Carlo technique. Within this analysis, we discuss the critical behavior of one-, two- and three-dimensional systems modeled by the first-neighbor spin-1 Ising model for different types of exchange interactions. The comparison of the results obtained according the Metropolis Monte Carlo simulations allows us to highlight the limits of the widely used mean-field theory approach. 
We also show, via a simple transformation, that for the special case where the bilinear and bicubic terms are set equal to zero in the Hamiltonian the partition function of the spin-1 Ising model can be reduced to that of the spin-1/2 Ising model with temperature dependent external field and temperature independent exchange
interaction times an exponential factor depending on 
the other terms of the Hamiltonian and confirm this result numerically by using the Metropolis Monte Carlo simulation.
Finally, we investigate the dependence of the critical temperature on the strength of long-range interactions included in the Ising Hamiltonian comparing it with that of the first-neighbor spin-1/2 Ising model.
\end{abstract}

\maketitle
\section{Introduction}
The classical spin-1/2  Ising model with nearest neighbor interactions for a lattice with $N$ sites was 
suggested by Lenz in 1920's is defined by the following Hamiltonian \cite{RefJ1} 
\begin{eqnarray}\label{E:HalfHamil}
 {\mathcal{H}_{\rm{spin}1/2}}
=-J\sum_{<ij>} s_{i}s_{j}-H\sum_{i=1}^{N}s_{i},
\end{eqnarray} 
where $s_{i}=+1$ or $-1$ is the spin variable. 
The notation $<ij>$ indicates a sum over nearest neighbors lattice sites, 
$J$ is the exchange 
constant which gives the interaction strength between two neighboring spins 
and $H$ is an external field applied to each degree
 of freedom. Ising solved the one-dimensional (1D) model in 1924 and,
 on the basis of the fact that 1D system had 
 no phase transition, he wrongly asserted that there was no phase 
transition in any dimension \cite{RefJ1,RefJ2}. Peierls
  proved that the model exhibits a phase transition in two or 
higher-dimensional lattices \cite{RefJ3}. 
The exact solution was found by 
  Onsager \cite{RefJ4,RefJ14}and Yang \cite{RefJ5} using 
algebraic approach and transfer matrix method. 
According to this analytical solution, in the
   two-dimensional (2D) square lattice a second-order phase transition at some 
critical temperature $T_c$ takes place when an external field $H
   $ tends to zero. This is  in agreement with the result of the series 
expansion method \cite{RefB1} and Monte Carlo simulations \cite{RefB2}. 
   Although there's no exact solution for three-dimensional (3D) lattices, 
it is possible to find the
critical temperature and critical exponents of the model using 
numerical methods like Monte Carlo simulations 
\cite{RefB2,RefJ6}.    

Spin-1/2  Ising model is appropriate to describe systems in which 
each degree of freedom has two states, but for systems with three 
states the spin-1 Ising model is more suitable \cite {RefB1}.
In the last decades, efforts have been made to study theoretically the underlying physics predicted by the spin-1 Ising model using mean-field theories and effective field theories \cite{RefnewJ1}. 
In particular, the mean-field solution in the presence of a random crystal field and the effects of the magnitude of the crystal field on the critical properties have been investigated. \cite{RefnewJ2,RefnewJ3,RefnewJ4,RefnewJ5}
  
On the other hand, the behavior of the tricritical point as a function of crystal-field interactions for honeycomb and its dependence on the strength of biquadratic and bilinear exchange interactions in square and
cubic lattices have been studied by using 
the effective-field theory \cite{RefnewJ6,RefnewJ7}.
   Recently, an analysis of spin-1 Ising model including only 
the bilinear term on tetrahedron recursive
lattices with arbitrary values of the coordination number 
has been performed to find an equation for the exact
determination of the critical points and all critical phases \cite{RefnewJ8}.
${\rm{He}^{3}-\rm{He}^{4}}$ mixture can be considered as spin-1 Ising model and some features of the mixture such as the $\lambda$ transition and the phase diagram of the mixture are described by the model 
 \cite{RefJ7}. However, despite their relevant results 
the above mentioned works focus on specific aspects of the critical properties of the spin-1 Ising model.
 
In this paper a systematic investigation of the critical properties exhibited by 1D, 2D (square) and 3D (cubic) systems modeled via the spin-1 classical Ising model Hamiltonian for different exchange interactions is performed overcoming some of the restrictions  
of the previous studies. 
The analytical investigation is carried out by using the low-temperature series expansion method applied to 
to the partition function. This is achieved after
determining the counts and the Boltzmann weights 
of the partition function depending 
on the full Hamiltonian of the classical spin-1 Ising model on 
both a square (2D) and on a cubic lattice (3D). 
 
The results obtained using the low-temperature series expansion
method are compared to the numerical ones determined via Metropolis
Monte Carlo simulations. The comparison of the exact Monte Carlo
results with the ones derived using the approximated mean-field theory
allows to highlight the limits of the latter approach that was widely
used in the past decades as mentioned above.

 The most general form of Hamiltonian of the spin-1 Ising model for a lattice with 
$N$ spin is  \cite{RefB1}
 \begin{eqnarray}\label{E:OneHamil}
{\mathcal{H}_{\rm{spin1}}}
=-J\sum_{<ij>}s_{i} s_{j} -K\sum_{<ij>}s_i^{2} s_j^{2}-D\sum_{i=1}^{N}s_{i}^{2} \nonumber \\
-L\sum_{<ij>}(s_{i}^{2} s_{j}+s_{i} s_{j}^{2} ) -H
\sum_{i=1}^{N}s_{i},     
\end{eqnarray}
where $s_{i}=+1$ or $0$ or $-1$, $K$ is the biquadratic coefficient, 
$D$ is the anisotropy coefficient and $L$ is the bi-cubic coefficient. 
Note that all coefficient appearing in (\ref {E:OneHamil}) 
have the dimension of an energy.
Sums are extended to the $N$ degrees of freedom and if the coefficients 
in (\ref {E:OneHamil}) are chosen to be positive, 
the ground state energy of the 
system corresponds to the configuration in which for all $i's$ we have $s_{i}=+1$. 
As the spin-1/2 model the 1D classical spin-1 Ising model does not exhibit 
any phase transition at finite temperatures.
In order to prove this claim we consider a 1D chain of spins with periodic 
boundary conditions. The free energy 
of the system, $F$ with entropy $S$ and energy $E$ at temperature 
$T$ by definition is given by \cite{RefB3}:
\begin{eqnarray}\label{E:FreeEnergy}
F=E-TS=E-k_{B}T\ln \Omega,                                                    
\end{eqnarray}
where $k_{B}$ is Boltzmann constant and $\Omega$ is the number 
of configurations with energy $E$.
Since we assume that all coefficients in (\ref {E:OneHamil})  are non-negative,
in the ground state all spin variables are up, i.e. for each site denoted 
by index $i$ we have $s_{i}=+1$ and 
the ground state configuration is $C=(+++...+++)$ where + means spin up. 
Hence, the energy and free energy of this 
configuration defined by $E_{0}$ and $F_{0}$ are   
\begin{eqnarray}\label{E:EnergyGs}
F_{0}=E_{0}=-N(H+D+J+K+2L).    
\end{eqnarray}

Now, by flipping one of the spins to $0$ or $-1$, the system will assume 
another configuration, which must have 
higher free energy with respect to ground state configuration $C$ in order to 
undertake a phase transition at $T\neq 0$. 
We denote these two possible configurations by 
$C_{1}=(++...++- ++...++)$ and $C_{1}^{'}=(++...++\ 0++...++)$ 
where $-$  and $0$ mean spin down and spin-less respectively.
The free energies, $F_{1}$ and $F_{1}^{'}$, associated 
with these configurations are
\begin{eqnarray}\label{E:EnergyC1}
F_{1}=E_{0}+4J+4L+2H-k_{B} T\ln N,                           
\end{eqnarray}       
\begin{eqnarray}\label{E:EnergyC'1}
F_{1}^{'}=E_{0}+2J+4L+H-k_{B} T\ln N.                             
\end{eqnarray}       
In both cases, in the thermodynamic limit and for $T\neq 0$ we get $F_{1},F_{1}^{'}<F_{0}$. 
Thus, the 1D version of spin-1 Ising model 
does not exhibit any phase transition at non-zero temperatures.  
However, for higher dimensions
the system described by the spin-1 Ising model exhibits a phase transition 
and, due to its enlarged parameter space,  
a much richer variety of critical behavior with respect to the spin-1/2 counterpart \cite{RefB1}. 
Since the study of the 2D and 3D spin-1 Ising 
 model  in its general form using either numerical or analytical tools 
becomes very interesting but at the 
same time very complicated, its critical
  behavior has only been studied in a few special cases in the literature. 
In one case just $J$ and $H$ are assumed 
  different from zero and the system described by 
 this spin-1 Ising model has been solved by applying 
low-temperature series expansion
  \cite{RefJ8,RefJ9}. In another case the spin-1 Ising model with 
  non-vanishing $J$, $D$, and $K$ has been investigated 
by using mean-field approximation 
   \cite{RefJ7}.  The latter one has been studied for $K=0$ 
by other methods like series expansion \cite{RefJ10}, 
renormalization group theory 
   \cite{RefJ11}, and Monte Carlo simulation \cite{RefJ12,RefJ13}. 

The paper is organized as follows. In section $2$ we shortly 
introduce the analytical and numerical methods 
applied to the spin-1 Ising model. In Section $3$ 
we firstly study the critical behavior
of the classical  spin-1 Ising model with nearest neighbor interaction
in 2D square and 3D cubic lattices, respectively comparing 
some critical features with those 
of the spin-1/2 Ising model, then last part of section $3$ 
deals with the long-range spin-1 Ising model  and 
the dependence of the critical temperature 
on the strength of long-range interactions.
 Finally, section $4$ is 
devoted to the conclusions. 

\section{Methods} 
 \label{sec:2}
In this section we briefly discuss the analytical and 
numerical approaches we have used to analyze  
the classical spin-1 Ising model. In the first and second 
subsections we describe the mean-field theory and the
low-temperature series expansion methods as representative 
analytical methods. 
The third subsection is a short 
description of the Metropolis Monte Carlo simulation which is a 
strong numerical tool that enables us 
to investigate the critical properties of the system.

\subsection{Mean-field theory}
\label{sec:2.1}
We use a systematic way of deriving the mean-field theory 
for some Hamiltonian $\mathcal{H}$
 in arbitrary dimension and coordinate number $z$ (cf. \cite {RefB1} 
where Mean-field theories are studied). 
 We begin from the Bogoliubov inequality
 \begin{eqnarray}\label{E:Bogo}
F\le \Phi=F_{0}+<\mathcal{H}-\mathcal{H}_{0}>_{0},  
 \end{eqnarray}
 
 where $F$ is the true free energy of the system, $\mathcal{H}_{0}$ 
is a trial Hamiltonian depending on 
 some variational parameters which we will introduce, $F_{0}$ is the 
 corresponding free energy, and $<...>_{0}$ denotes an 
average taken in the ensemble defined by 
 $\mathcal{H}_{0}$. The mean-field free energy, $F_{\rm{MF}}$, 
is then defined by minimizing $\Phi$ with respect 
 to the variational parameters. In order to see how this method works let us consider the
 most general form of the classical spin-1 Ising model given by (\ref {E:OneHamil}). 
 We introduce the trial Hamiltonian as the following
 \begin{eqnarray}\label{E:TrialH}
\mathcal{H}_{0}=-H_{0} \sum_{i=1}^{N}s_{i} -D_{0} \sum_{i=1}^{N}s_{i}^{2}. 
 \end{eqnarray}
 
Hence, there are two variational parameters, $H_0$ and $D_0$, 
which will be determined by minimizing the functional $\Phi$.
Assuming that the lattice is translationally invariant 
it is straightforward to find the partition function 
and the mean-field free energy. Consequently one 
 can easily compute the magnetization, $M=<s_{i}>_{0}$ 
defined in dimensionless units as the thermal
average of the spin variable and the thermal average of the
 square of the spin variable $\tau=<s_i^2>_0$ as follows
 \begin{eqnarray}\label{E:gen_M_mf}
M=\frac{2e^{\beta(D+LzM+Kz\tau)} \sinh[\beta(JzM+H+Lz\tau)]}{1+2e^{\beta(D+LzM+Kz\tau)} \cosh[\beta(JzM+H+Lz\tau)]}, \nonumber \\ 
\end{eqnarray}  
\begin{eqnarray}\label{E:gen_sigma_mf}
\tau=\frac{2e^{\beta(D+LzM+Kz\tau)} \cosh[\beta(JzM+H+Lz\tau)]}{1+2e^{\beta(D+LzM+Kz\tau)} \cosh[\beta(JzM+H+Lz\tau)]}. \nonumber \\
\end{eqnarray} 

Solving two self consistent equations (\ref {E:gen_M_mf}) and (\ref {E:gen_sigma_mf}) 
simultaneously, one can find the mean-field
 magnetization as a function of the temperature.
 
We remind that, although the use of 
 mean-field theory gives us some valuable information about the behavior of the system 
 and specifically it allows us to reproduce the phase diagram of the model in 
a relatively simple and qualitative way, 
from a quantitative point of view
 the results are not generally exact  for the case of 2D and 3D lattices so that 
in this case we need to use more precise analytical and numerical methods
allowing us to have a more exact  knowledge of the critical behavior of these
systems also quantitatively.

\subsection{low-temperature series expansion method}
\label{sec:2.2}

Another possible way to investigate the spin-1 Ising model
is to use low-temperature series expansion. The idea is to start from a 
completely ordered configuration, i.e. the ground state, 
and then flip spins one by one, and take all 
the configurations into account to compute the partition function, $Z$ 
 as the following: \cite{RefB3} 
\begin{eqnarray}\label{E:PartLTSEformula}
Z=e^{-\frac{E_{0}}{k_{B} T}} (1+\sum_{n=1}^{\infty}\Delta Z_{N}^{(n)}),                                          
\end{eqnarray}
where $E_0$ is ground state energy, and 
$\Delta Z_{N}^{(n)}$ is the sum of Boltzmann factors with energy 
that is measured with respect to the ground state energy 
when $n$ spins are flipped starting from the ground state configuration. 
Two factors contribute to the Boltzmann factor $\Delta Z_{N}^{(n)}$, namely
the number of ways of flipping $n$ spins with a specific Boltzmann weight, 
or counts, and the corresponding 
Boltzmann weights \cite{RefB3}.
\begin{table}
\caption{The counts and the Boltzmann weights contributing to the to the low-temperature series 
expansion of the partition function of a square lattice classical spin-1 Ising model for different numbners of flipped spins ($N_f$).}
\label{tab:1}       

\begin{tabular}{|c|c|c|}
\hline
$N_f$ & Count & Boltzmann weight \\
\hline

$1$ & $N$ & $x^{8}y^{2}z^{8}$ \\
$1$ & $N$ & $x^{4}yz^{8}uw^{4}$  \\
$2$ & $2N$ & $x^{12}y^{4}z^{16}$  \\
$2$ & $4N$ & $x^{10}y^{3}z^{14}uw^{4}$   \\
$2$ & $2N$ & $x^{7}y^{2}z^{14}u^{2}w^{7}$  \\
$2$ & $N(N-5)/2$ & $x^{16}y^{4}z^{16}$  \\
$2$ & $N(N-5)$ & $x^{12}y^{3}z^{16}uw^{4}$  \\
$2$ & $N(N-5)/2$ & $x^{8}y^{2}z^{16}u^{2}w^{8}$   \\
$3$ & $2N$ & $x^{16}y^{6}z^{24}$   \\
$3$ & $4N$ & $x^{14}y^{5}z^{22}uw^{4}$   \\
$3$ & $2N$ & $x^{16}y^{5}z^{20}uw^{4}$   \\
$3$ & $4N$ & $x^{13}y^{4}z^{20}u^{2}w^{7}$   \\
$3$ & $2N$ & $x^{12}y^{4}z^{20}u^{2}w^{8}$   \\
$3$ & $2N$ & $x^{10}y^{3}z^{20}u^{3}w^{10}$   \\
$3$ & $4N$ & $x^{16}y^{6}z^{24}$   \\
$3$ & $8N$ & $x^{14}y^{5}z^{22}uw^{4}$   \\
$3$ & $4N$ & $x^{16}y^{5}z^{20}uw^{4}$   \\
$3$ & $8N$ & $x^{13}y^{4}z^{20}u^{2}w^{7}$   \\
$3$ & $4N$ & $x^{12}y^{4}z^{20}u^{2}w^{8}$   \\
$3$ & $4N$ & $x^{10}y^{3}z^{20}u^{3}w^{10}$   \\
$3$ & $2N(N-8)$ & $x^{20}y^{6}z^{24}$   \\
$3$ & $2N(N-8)$ & $x^{16}y^{5}z^{24}uw^{4}$   \\
$3$ & $4N(N-8)$ & $x^{18}y^{5}z^{22}uw^{4}$   \\
$3$ & $2N(N-8)$ & $x^{15}y^{4}z^{22}u^{2}w^{7}$   \\
$3$ & $4N(N-8)$ & $x^{14}y^{4}z^{22}u^{2}w^{8}$   \\
$3$ & $2N(N-8)$ & $x^{11}y^{3}z^{22}u^{3}w^{11}$   \\
$3$ & $N(N^{2}-15N+62)/6$ & $x^{24}y^{6}z^{24}$   \\
$3$ & $N(N^{2}-15N+62)/2$ & $x^{20}y^{5}z^{24}uw^{4}$   \\
$3$ & $N(N^{2}-15N+62)/2$ & $x^{16}y^{4}z^{24}u^{2}w^{8}$   \\
$3$ & $N(N^{2}-15N+62)/6$ & $x^{12}y^{3}z^{24}u^{3}w^{12}$   \\
$4$ & $N$ & $x^{12}y^{4}z^{24}u^{4}w^{12}$   \\
$4$ & $18N$ & $x^{13}y^{4}z^{26}u^{4}w^{13}$   \\

\hline
\end{tabular}\\

\end{table}
\normalsize
We consider the spin-1 Ising model described by the Hamiltonian (\ref {E:OneHamil}) for a 
square lattice as an example.
For the sake of simplicity, we introduce the following parameters
\begin{eqnarray}\label{E:Variables}
x=e^{-\beta J},y=e^{-\beta H},z=e^{-\beta L},u=e^{-\beta D},w=e^{-\beta K}, \nonumber \\          
\end{eqnarray} where $\beta =\frac{1}{k_{B}T}$. Then we need to calculate 
the counts for different configurations. 
The computation of the counts is more complicated with respect 
to that of the spin-1/2 Ising model  because
in the spin-1 Ising model each spin can choose among three possible states. 
Table \ref {tab:1} shows Boltzmann weights 
 and their counts for some configurations in which a few spins are flipped. 
Starting from these calculations summarized in Table \ref {tab:1} we 
finally obtain the low-temperature 
 series expansion of the partition function for the spin-1 Ising model :
\begin{align}
Z_{\rm{spin1}}^{\rm{square}} = &  e^{-\beta E_{0}} \bigg(1+Nx^{4} yz^{8} uw^{4}+2Nx^{7} y^{2} z^{14} u^{2} w^{7}+ 
\nonumber\\ & Nx^{8} y^{2} z^{8}  
        -\frac {5N}{2} x^{8} y^{2} z^{16} u^{2} w^{8}+ 4Nx^{10}y^{3} z^{14} uw^{4}+ \nonumber \\ &6Nx^{10} y^{3} z^{20} u^{3} w^{10}
        -16Nx^{11} y^{3} z^{22} u^{3} w^{11}+ \nonumber \\&2Nx^{12} y^{4} z^{16}-5Nx^{12} y^{3} z^{16} uw^{4}  
        +\nonumber \\&6Nx^{12} y^{4} z^{20} u^{2} w^{8} +\frac {31N}{3} x^{12} y^{3} z^{24} u^{3} w^{12}+\nonumber \\&Nx^{12} y^{4} z^{24} u^{4} w^{12}
        +12Nx^{13} y^{4} z^{20}u^{2} w^{7}+\nonumber \\&18Nx^{13} y^{4} z^{26} u^{4} w^{13}+O(x^{14})\bigg) \label{E:PartLTSE}
\end{align}

We will follow the same procedure
by finding the Boltzmann weights and the associated counts for the 3D cubic lattice
in one of the special cases outlined in section $3$. 
\subsection{Metropolis Monte Carlo method}
\label{sec:2.3}

Monte Carlo method is a powerful numerical tool widely used 
to evaluate discrete spin models
 like Ising models to investigate the behavior of the associated 
thermodynamic functions of the models. 
 It is also very popular to study continuous spin systems like XY and 
Heisenberg model, fluids, polymers, 
 disordered materials, and lattice gauge theories (cf. \cite {RefB4} 
where Monte Carlo Methods are studied). 
In this paper we use Metropolis Monte Carlo simulation. 
The algorithm can be summarized in three steps:
 
1. Set up of the lattice sites.  
To do it, for example in the case of 2D square lattice, we define a 2D array
 with $l_{x} \times l_{x}$ spin, which is called spin$[i][j]$, 
where $i$ and $j$ determine a specific lattice site.
 
2. Initialization of the system.
We use a function named
init $(l_x$, $J$, $K$, $D$, $L$, $H)$
to set the initial state of the system, which in principal can be 
chosen arbitrarily. In the simulations we choose 
a completely ordered state as the initial state with 
maximum magnetization. In this function, $l_{x}$ is the 
number of spins in each direction, and $J$, $K$, $D$, $L$, 
and $H$ are the values of the coefficients in the 
Hamiltonian of the spin-1 Ising model 
given by equation (\ref {E:OneHamil}).
	
3. Use of a main loop in the main program 
to update the system many times. 
The function mc$(T)$ takes the 
temperature $T$ and uses Metropolis update. 
This function at first chooses randomly one of the spins in the 
lattice and flips that spin. 
For instance, if the randomly chosen spin is $+1$ 
it is flipped to $0$ or $-1$ with the same
probability. 
The probability that the system is allowed to move 
from the initial state to the final state is
$$
P(\mathrm{initial} \to \mathrm{final})=
\begin{cases}
1, \ \ \text{if} \ \ E_{\mathrm{final}}<E_{\mathrm{initial}}\\
e ^{ ({-\beta (E_{\mathrm{final}}-E_{\mathrm{initial}}}))}, \ \ \ \text{otherwise} \\
\end{cases}
$$
By updating the system a sufficient number of times, 
it eventually reaches the equilibrium state at any temperature. 
Finally, it is possible to determine the thermodynamic functions 
such as magnetization, and susceptibility using 
following formulas \cite {RefB2}:
\begin{eqnarray}\label{E:MagMC}
M=\frac {1}{N} \sum_{i=1}^{N}s_{i}, 
\end{eqnarray}
\begin{eqnarray}\label{E:SuscMC}
\chi=\frac {1}{k_{B} T} (<M^{2}>-<M>^2 ).   
\end{eqnarray}
\section{Results and discussions: special cases of spin-1 Ising 
Hamiltonian for 2D square and 3D cubic lattices}
\label{sec:3}

Owing to the previous arguments, in principle we know how 
to calculate the partition function associated with 
(\ref {E:OneHamil}) expressing the Hamiltonian of 
the spin-1 Ising model and consequently 
we can characterize thermodynamically
the model. However, since the general form of the Hamiltonian
 is very complex and the number of parameters 
appearing in the parameter space is high, 
it is not possible to study analytically and/or numerically 
in an exact way the critical behavior of the full Hamiltonian. 
It is thus useful to understand better 
the critical behavior of the spin-1 Ising model 
focusing our attention on 
some special cases with a lower number of parameters 
that are numerically solvable. These cases 
will be discussed in the following subsections. 
In the last subsection we will introduce the long-range 
spin-1 Ising model in analogy with the well-known 
spin-1/2 model to find out 
how its critical temperature depends on the
magnitude of the long-range interaction. 
\subsection{Case 1}
\label{sec:3.1}
In this subsection, we restrict ourselves to the specific case in which all 
the coefficients in (\ref {E:OneHamil})  are 
zero except $J$ and $H$, so that the Hamiltonian is given by
\begin{eqnarray}\label{E:Hamil1}
{\mathcal{H}_{\rm{Ising1}}^{(1)}}
=-J\sum_{<ij>}s_{i} s_{j} -H\sum_{i=1}^{N}s_{i}.    
\end{eqnarray}
In this special case, the series expansion of the partition function for a 2D square lattice  
is obtained by setting $D$, $K$, and $L$ zero in
(\ref {E:PartLTSE}). 
Let's write down the free energy $F$, the magnetization $M$, 
and the susceptibility $\chi$ as follows \cite {RefB3}
\begin{eqnarray}\label{E:Fener} 
F=-k_{B}T\ln Z, 
\end{eqnarray}
\begin{eqnarray}\label{E:Mag} 
M=-\frac {1}{N}\lim_{H \to 0} \left (\frac {\partial F}{\partial H} \right) _{T},  
\end{eqnarray}
\begin{eqnarray}\label{E:Susc} 
\chi=\frac {1}{N\beta}\lim_{H \to 0} \left (\frac {\partial^2 \ln Z}{\partial H^2} \right) _{T}.  
\end{eqnarray}

Now we can write down the series expansion of $M$, and $\chi$
for the square lattice:
\begin{align}\label{E:MagLSR} 
M_{\rm{square}}^{(1)}=&1-x^{4}-4x^{7}+3x^{8}-30x^{10}+48x^{11}-52x^{12} \nonumber \\ &-120x^{13}+O(x^{14} ),
\end{align}  
\begin{align}\label{E:SuscLSR} 
\chi_{\rm{square}}^{(1)}=&\beta(x^{4}+8x^{7}-6x^{8}+90x^{10}-144x^{11}+192x^{12} \nonumber \\ &+480x^{13}+O(x^{14} )).
\end{align} 

Likewise, one can find corresponding expressions for a 3D lattice. 
For simplicity we take a simple cubic lattice with 
each site having $z=6$ nearest neighbors. Since the 
procedure is similar to what we have done for 
square lattice we only write down the final 
expression of the above quantities as follows:
\begin{align}\label{E:PqrtCub1} 
Z^{(1)}_{\rm{cube}}&= e^{-\frac {E_0}{k_B T}} \bigg[1+Nx^6 y+3Nx^{11} y^2 \nonumber\\ &+(\frac {N(N-7)}{2}+N) x^{12} y^2 
    +21Nx^{16} y^3\nonumber\\ &+3N(N-12) x^{17} y^3 
    \nonumber\\
    &+\bigg(N(N-7)+ \frac {N(N^2-3N+2)}{6} \nonumber \\ &-3N(N-12)-15N\bigg) x^{18} y^3 \nonumber\\
    &+21Nx^{20} y^4+77Nx^{21} y^4 \nonumber \\ & +\bigg(3N(N-17)+12N(N-16) \nonumber \\ &+\frac{3N(N-17)}{2} +3N(N-20)+6N(N-12)\bigg) x^{22} y^4 \nonumber \\ & +O(x^{23} )\bigg], 
\end{align}
 
\begin{align}\label{E:MagCub1}
M_{\rm{cube}}^{(1)}=&1-x^6-6x^{11}+5x^{12}-63x^{16}+108x^{17}-43x^{18}\nonumber\\ &-84x^{20}-308x^{21} 
     +1602x^{22}+O(x^{23} ),  
\end{align} 
\begin{align}\label{E:SuscCub1} 
\chi_{\rm{cube}}^{(1)}=&\beta[x^6+12x^{11}-10x^{12}+189x^{16}-324x^{17}+129x^{18}\nonumber\\&+336x^{20}
     +1232x^{21}-6408x^{22}+O(x^{23} )].  
\end{align} 

Equations (\ref {E:MagLSR}) and (\ref {E:SuscLSR}) obtained from 
combinatorial combinations are in agreement with the results found in
\cite {RefJ8} using finite lattice method for a 2D square lattice. 
\begin{figure}
\begin{center}
\includegraphics[width=0.45\textwidth]{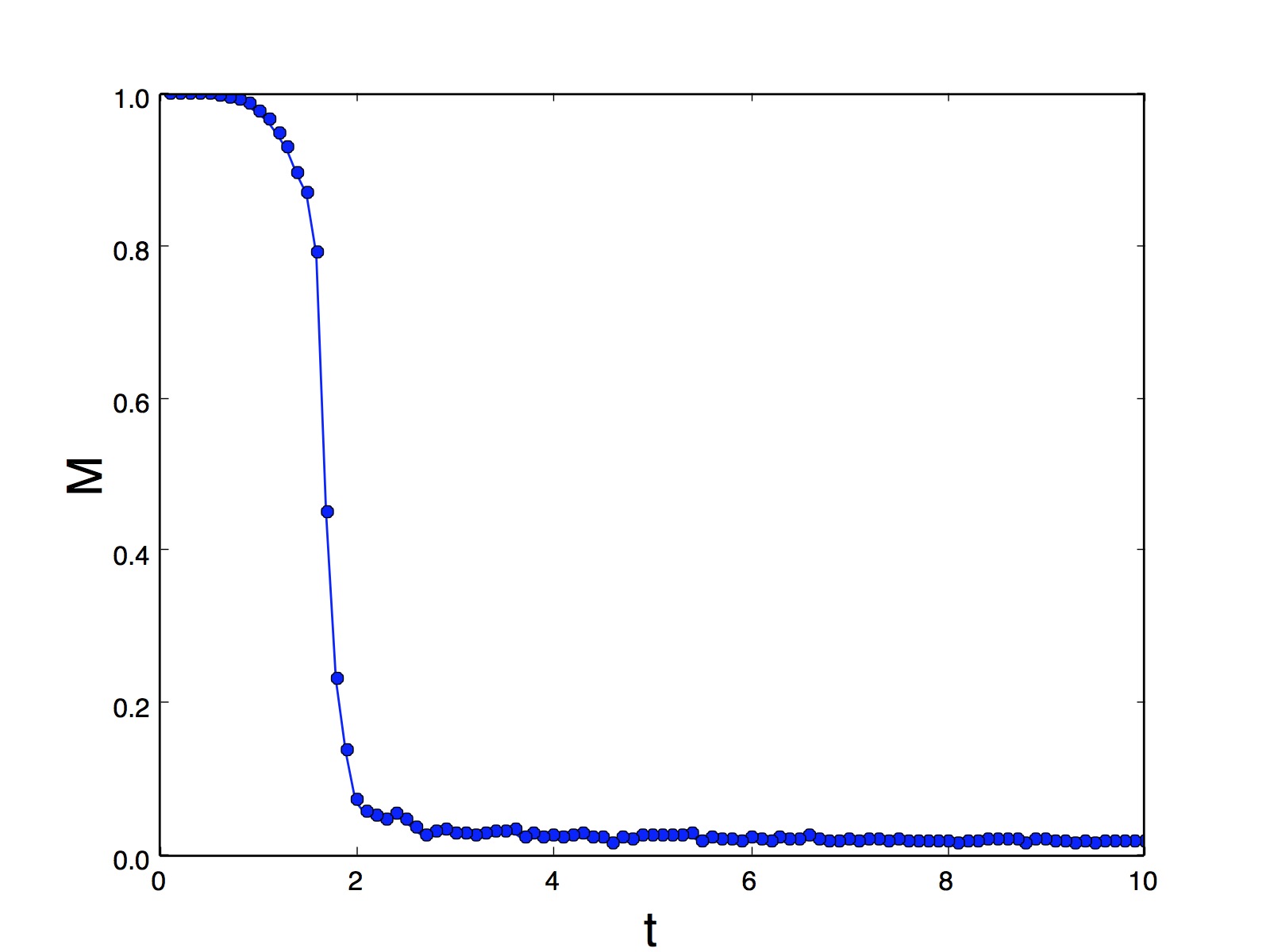}
\caption{\label{fig:MT-1}Magnetization $M$ vs. reduced temperature 
$t =\frac {k_{B}T}{J}$ obtained by using Metropolis Monte Carlo simulation for a square lattice with $40\times40$ sites for the 
case 1. The error bars are smaller than size of the markers.}
\end{center}
\end{figure}
As shown by Enting, Guttmann and Jensenin \cite {RefJ8} at least the first 60 
terms of low-temperature series expansion of thermodynamic functions are needed 
to have a physically consistent result. 
 The critical temperature has been finally approximated as follows
\begin{eqnarray}\label{E:CriticalTemp1}
\exp({-\frac {J}{k_{B}T_{c, SE}^{(1, \rm{square})}}})=x_{c, SE}^{(1, \rm{square})}=0.554075\pm 0.000015. \nonumber \\
\end{eqnarray}                                   
Also the calculation of the critical exponents $\beta$ and $\gamma$ 
associated to $M$ and $\chi$, respectively 
lead to the conclusion that the model belongs
to the same universality class as spin-1/2 Ising model. 
Now, we apply the Metropolis Monte Carlo simulation on a 
 2D square lattice and examine these conclusions.
 Figure \ref{fig:MT-1} shows the magnetization 
versus the reduced temperature, $t =\frac {k_{B}T}{J}$  for a square 
lattice with  $40\times40$  spins. 
\begin{figure}
\begin{center}
  \includegraphics[width=0.45\textwidth]{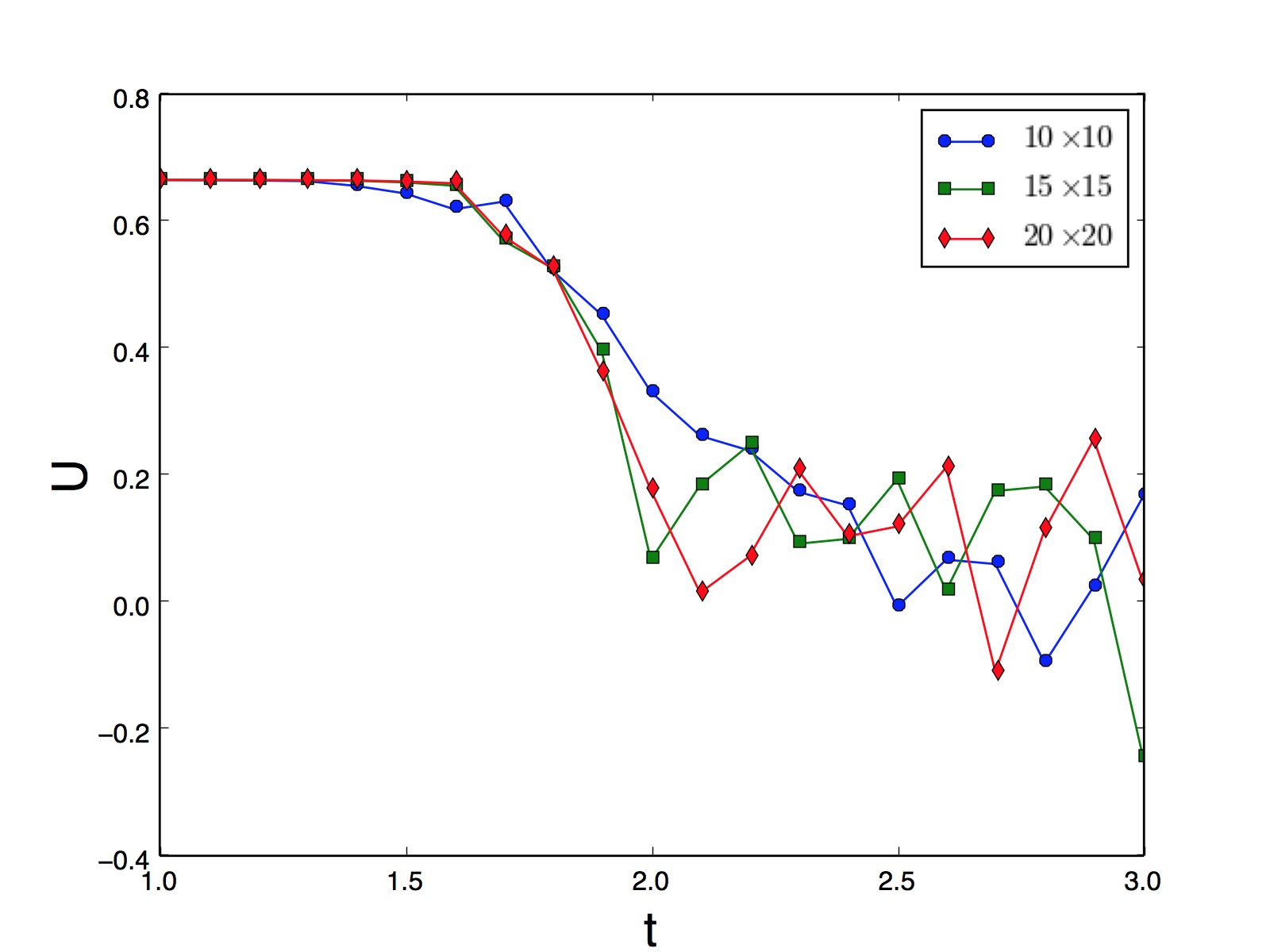}
\caption{Binder Cumulant vs. $t =\frac {k_{B}T}{J}$ 
for different size lattices for Case 1. The error bars are smaller than size of the markers.}
\label{fig:TC-1}       
\end{center}
\end{figure}
It shows that in the high temperature regime, system 
is in a disordered phase and the 
magnetization is zero, and for a reduced temperature 
$t=\frac{k_BT}{J}\simeq1.7$ a critical phase transition occurs and the 
system evolves towards the 
ordered phase. Finally, for very low-temperatures the magnetization 
is close to one as expected. 

The Binder cumulant defined as
\begin{eqnarray}\label{E:BC}
U=1-\frac{<M^4 >}{3<M^2 >^2 },      
\end{eqnarray}
 is an observational tool to estimate critical points.
It turns out that the intersection of $U-T$ curves, for networks 
with different number of sites, gives the critical temperature 
of the lattice with a good accuracy \cite {RefB4,RefJ20}. 
Figure \ref{fig:TC-1} shows how we can use the Binder cumulant to determine the 
critical temperature considering three different size lattices. 
In this figure the Binder cumulants for three 2D square lattices 
with $5\times5$, $10\times10$, and $15\times15$ sites are displayed. 
The intersection of the curves corresponds to the 
critical point given by
\begin{eqnarray}\label{E:TC1}
\frac {k_{B}T_{c,\rm{MC}}^{(1,\rm{square})}}{J}=1.70\pm0.01.    
\end{eqnarray} 
This result is in the accordance with the critical 
temperature found from series expansion method 
given by (\ref{E:CriticalTemp1}). 

In order to complete our discussion about this specific case we shall 
find some of the critical exponents of the model using the data 
we have obtained from simulation by the so called finite lattice 
method \cite {RefB2}. Let us firstly evaluate the $\beta$ critical exponent. 
In order to calculate the $\beta$ exponent what is usually 
done is to plot $y=\ln M$ vs. $x=\ln l_{x}$. The slope 
of this graph is the $\beta$ exponent. 
Likewise the slope of $y=\ln \chi$ vs. $x=\ln l_{x}$ 
gives the $\gamma$ exponent. 
We eventually
find 
\begin{eqnarray}\label{E:Beta1}
\beta = 0.13\pm0.01,    
\end{eqnarray} 
\begin{eqnarray}\label{E:gamma1}
\gamma = 1.78\pm0.05.     
\end{eqnarray}     

These observations suggest that spin-1 Ising model 
governed by Hamiltonian (\ref {E:Hamil1})  
belongs to the same universality class of the spin-1/2 
Ising model in agreement with series expansion method. 
In order to have an approximation of the critical temperature 
for a 3D cubic lattice we 
use again the Binder cumulant.
 The critical temperature is:
\begin{eqnarray}\label{E:TC1-cube}
\frac {k_{B}T_{c,\rm{MC}}^{(1,\rm{cube})}}{J}=3.2\pm0.1 .    
\end{eqnarray}

\subsection{Case 2}
\label{sec:3.2}    
In this subsection, we consider the Hamiltonian of the spin-1 
Ising model given by (\ref {E:OneHamil}) and 
we assume $J=L=0$.                                                         
Therefore, the Hamiltonian of the system is:
\begin{eqnarray}\label{E:Hamil2}
{\mathcal{H}_{\rm{Ising1}}^{(2)}}
=-K\sum_{<ij>}s_{i}^{2} s_{j}^{2} -D\sum_{i=1}^{N}s_{i}^{2} -H\sum_{i=1}^{N}s_{i},                           
\end{eqnarray}  
where $s_{i}=0$,$+1$,$-1$.

The Hamiltonian in the absence of an external magnetic field  
has been studied by Griffiths \cite {RefJ15}, who showed that 
the statistical mechanics of the spin-1 Ising model 
can be reduced to that of spin-1/2 Ising model.
Although it would be possible in principle to solve this model 
with $H\ne0$ using series expansion method, 
in this section we outline a very 
simple analytical solution not directly obtained with series expansion and 
we compare it with the results derived by means of 
Metropolis Monte Carlo technique simulations. 
We will prove that, by applying a simple transformation, 
the spin-1 Ising model is reduced to the spin-1/2 Ising model, 
with a constant exchange, but a temperature dependent 
external field \cite{wu1978phase}. Before starting our discussion about 
the Hamiltonian (\ref {E:Hamil2}) we make a very simple consideration. 
If we assume that $K=0$ the system does not exhibit any 
critical behavior because of the lack of 
a collective behavior in the system. Therefore, all the arguments 
in this subsection are valid only for $K\ne0$ that marks
the collective behavior in this case. 

We consider a lattice of arbitrary dimensions and coordinate number $z$ 
described by the Hamiltonian (\ref {E:Hamil2}).
The partition function, $Z$ is defined by \cite {RefB1}
\begin{eqnarray}\label{E:PartDef}
Z_{\rm{spin1}}=\sum_{s=0,+1,-1}e^{-\beta \mathcal{H}}                                               
\end{eqnarray} 
Substituting (\ref {E:Hamil2}) in (\ref {E:PartDef}) and using the 
transformation $t_{i}=2s_{i}^{2}-1$ and some straightforward 
calculations we get
\begin{eqnarray}\label{E:Part2}
Z_{\rm{spin1}}^{(2)}=e^{N\beta C} \sum_{t_i=+1,-1}e^{R\sum_{i=1}^{N}t_{i} +Q\sum_{<ij>}t_{i} t_{j} },      
\end{eqnarray} 
where
\begin{eqnarray}\label{E:CDef}
C=\frac {1}{2\beta}  \ln(2\cosh \beta H)+\frac{D}{2}+\frac{Kz}{8},
\end{eqnarray}
\begin{eqnarray}\label{E:RDef}
R=\frac{1}{2}  \ln(2\cosh \beta H)+\frac {\beta D}{2}+\frac {\beta Kz}{4}
\end{eqnarray}
\begin{eqnarray}\label{E:QDef}
Q=\frac {\beta K}{4}.
\end{eqnarray}
Thus, according to the equation (\ref {E:Part2}) the partition function 
of the spin-1 Ising model given by the Hamiltonian 
equation (\ref {E:Hamil2}) with an appropriate transformation is reduced 
to that of the spin-1/2 Ising model with temperature dependent
 external field $\frac {R}{\beta}$ and temperature independent exchange 
interaction $\frac {Q}{\beta}$ times an exponential factor:
\begin{eqnarray}\label{E:PartitionDef}
Z_{\rm{spin1}}^{(2)}=e^{N\beta C}\times Z_{\rm{spin1/2}} (\frac {R}{\beta},\frac {Q}{\beta}).
\end{eqnarray}

The Hamiltonian of the equivalent spin-1/2 Ising model 
with external field $\frac {R}{\beta}$ and exchange coefficient 
 $\frac {Q}{\beta}$ is
\begin{eqnarray}\label{E:eqHamil}
\mathcal{H}_{\rm{spin1/2}}
 =-\frac {R}{\beta} \sum_{i=1}^{N}t_{i} -\frac {Q}{\beta} \sum_{<ij>}t_{i} t_{j}   
 \end{eqnarray}                                    
 So (\ref{E:Part2}) and (\ref {E:eqHamil})  show that spin-1 model 
with Hamiltonian (\ref {E:Hamil2}) at 
 temperatures less than the critical temperature exhibits 
the first order phase transition by crossing following surface:
 \begin{eqnarray}\label{E:PhTr2}
 \frac {R}{\beta}=0.
  \end{eqnarray}
  \begin{figure}
  \begin{center}
  \includegraphics[width=0.45\textwidth]{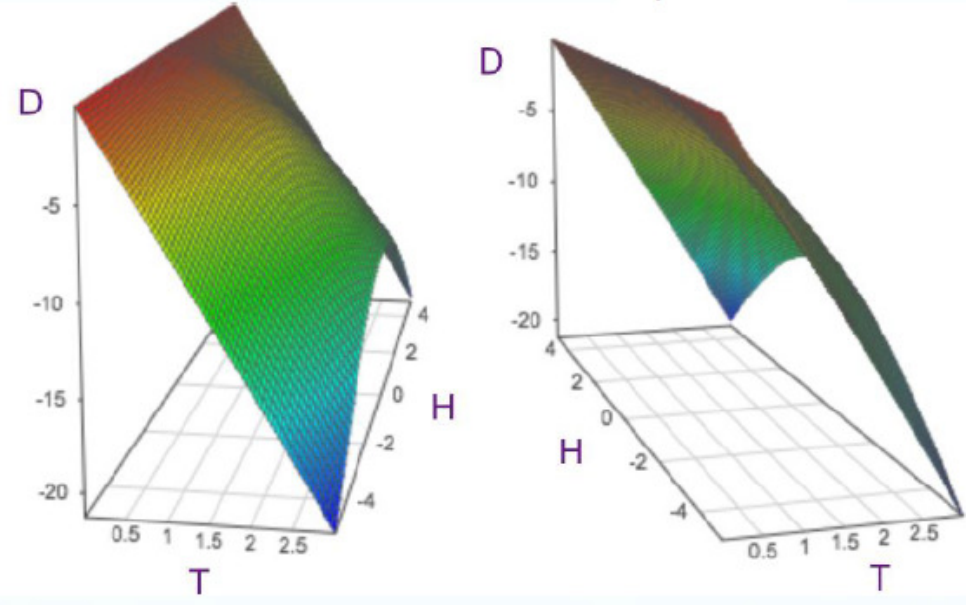}
\caption{The surface $\frac {R}{\beta }=0$ in the $(T, D, H)$ space 
for $K=1$ in two views. We assume $k_{B}=1$
for simplicity.} 
\label{fig:Phase-2}       
\end{center}
\end{figure}

 Figure \ref {fig:Phase-2} displays the phase transition 
surface of the spin-1 
Ising model governed by Hamiltonian (\ref {E:Hamil2}) for $K=1$. 
So if we fix the temperature to a low enough value that is less 
than the critical temperature and change  
 the external field $H$  in a wide enough range for appropriate 
values of $D$, there are two first-order 
 phase transitions. 
 Furthermore, it is clear that there is a maximum value of $D$ up 
to which the phase transition is possible; 
 one can find this maximum value in the limit $T\to0$, and $H\to0$:
  \begin{eqnarray}\label{E:Dmax}
 D_{\rm{max}}=-\frac {Kz}{2}.
  \end{eqnarray}
 Since we know the exact critical temperature of the  spin-1/2  
Ising model \cite {RefJ4}
 for the 2D square lattice,  we can easily 
 find the critical temperature for case 2 for such a lattice. 
The critical temperature of the 2D square lattice 
 spin-1/2  Ising model with Hamiltonian (\ref {E:eqHamil}) equation 
or equivalently for the
spin-1 Ising model  described  by (\ref {E:Hamil2}) is
 \begin{eqnarray}\label{E:TC2}
 T_{c}^{(2,\rm{square})}=\frac {K}{2k_{B}\ln(1+\sqrt{2})}  
  \end{eqnarray}
\begin{figure}
\begin{center}
 \includegraphics[width=0.45\textwidth]{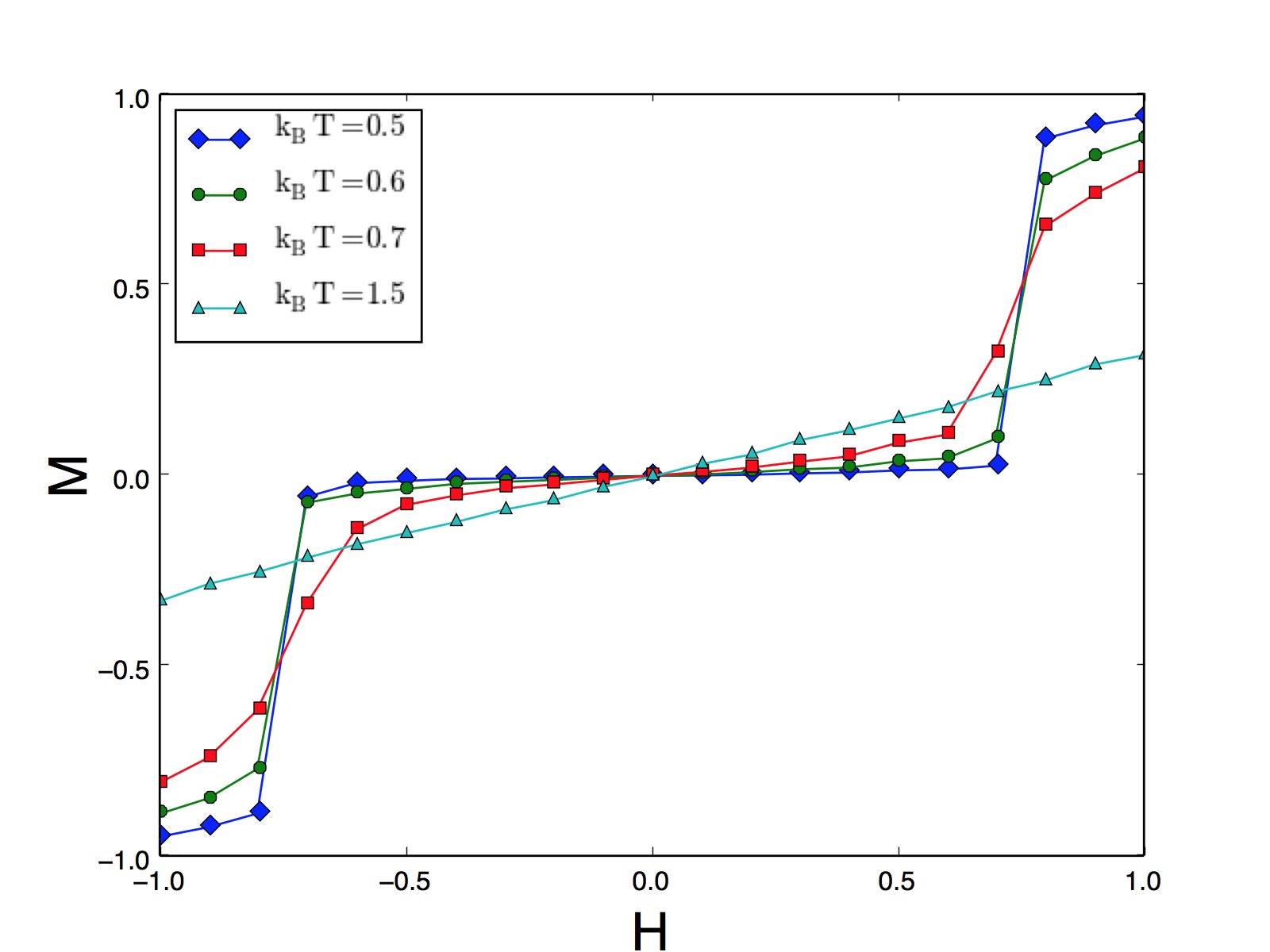}
\caption{$M-H$ curves for case 2 with $D=-2.8$, and $K=1$
at different temperatures: $k_BT=0.5$, $k_BT=0.6$, $k_BT=0.7$, $k_BT=1.5$, 
for a $40\times40$ square lattice obtained from Metropolis 
Monte Carlo simulation. The error bars are smaller than size of the markers.} 
\label{fig:MH-2}       
\end{center}
\end{figure}
 We now perform Metropolis Monte Carlo simulation for 2D square 
lattice to compare the simulation 
results with the analytical calculation with special regard to 
the critical temperature. 
We set $K=1$, and use the suitable value for $D$, and change 
$H$, from $+1$ to $-1$ for 
 different values of $T$. In Figure \ref {fig:MH-2} the results 
for a $40\times40$ 2D lattice are shown. 
The magnetization $M$ is plotted versus the external field, $H$ 
for different values of temperature. 
 As it can be seen, for very low $T$, there are two jumps in the 
curve. One jump corresponds to a positive value of 
 $H$, in which the magnetization jumps from $1$ to zero. 
Another jump occurs at a negative value of $H$, 
 and in this case magnetization has a sudden variation from zero to $-1$. 
This result is conceptually is in 
 accordance with our analytical solution. In fact, we have two first order phase 
transitions corresponding to the two 
 values of $H$, where coefficient $\frac {R}{\beta}$ is zero. Moreover,
 the critical temperature can be estimated
 by plotting the susceptibility versus temperature; it turns out that
 \begin{eqnarray}\label{E:MCTC2}
 \frac{k_BT_{c,\rm{MC}}^{(2,\rm{square})}}{K}=0.57\pm0.01
 \end{eqnarray} which agrees with the expression of the analytical 
critical temperature given by (\ref {E:TC2}). 
 The qualitative behavior of the 3D lattice system is very similar 
to the square lattice as we expect 
 and the critical phase transition occurs at 
 \begin{eqnarray}\label{E:MCTC2-cube}
 \frac{k_BT_{c,\rm{MC}}^{(2,\rm{cube})}}{K}=1.20\pm0.1.
 \end{eqnarray}
 
\subsection{Case 3}
\label{sec:3.3}    
In this section we consider the Hamiltonian (\ref {E:OneHamil}) with $K=0$:
\begin{align}\label{E:Hamil3}
{\mathcal{H}_{\rm{Ising1}}^{(3)}}
=&-J\sum_{<ij>}s_{i} s_{j} -D\sum_{i=1}^{N}s_{i}^{2} -L\sum_{<ij>}(s_{i}^{2} s_{j}+s_{i} s_{j}^{2} ) \nonumber \\ &-H
\sum_{i=1}^{N}s_{i},
 \end{align}
 and we apply mean-field approximation to obtain some physical 
quantities characterizing the model. 
More specifically, we get $F_{\rm{MF}}$ and, as a result, two 
self-consistent equations for mean-field magnetization $M$ and 
the thermal average of the square of the spin variable $\tau$:
\begin{align}\label{E:F_mf}
F_{\rm{MF}}^{(3)}=&-Nk_B T \times \nonumber \\ &\ln{\bigg \{1+2e^{\beta(D+LzM)}   \cosh[\beta(zJM+H+Lz\tau)] \bigg \}} \nonumber \\ & +\frac {NJz}{2} M^2+NLzM\tau,       
\end{align}
\begin{eqnarray}\label{E:M_mf}
M=\frac{2e^{\beta(D+LzM)} \sinh[\beta(JzM+H+Lz\tau)]}{1+2e^{\beta(D+LzM)} \cosh[\beta(JzM+H+Lz\tau)]},  \nonumber \\
\end{eqnarray}  
\begin{eqnarray}\label{E:sigma_mf}
\tau=\frac{2e^{\beta(D+LzM)} \cosh[\beta(JzM+H+Lz\tau)]}{1+2e^{\beta(D+LzM)} \cosh[\beta(JzM+H+Lz\tau)]}. \nonumber \\
\end{eqnarray} 

One usual way to solve a system of nonlinear and 
transcendental algebraic equations 
like (\ref {E:M_mf}) and (\ref {E:sigma_mf}) is Newton's method
\cite {RefB5}.
Eq.s (\ref {E:M_mf}) and (\ref {E:sigma_mf}) imply that no phase transitions are predicted by the mean-field 
theory for nonzero $L$. 
For instance, let us assume that in Hamiltonian (\ref {E:Hamil3}) 
$L$ is positive and all the other coefficients vanish.
When $T$ decreases, the system tends towards a configuration
 in which all the spins are up; for negative 
$L$ the system instead tends to a configuration with all spins down.
 Hence, $L$ plays a role similar to $H$.
Furthermore, as should be expected, at high temperatures the system is 
in the disordered phase. In other words, the number of 
spin up, spin down and spin-less sites are equal, and 
consequently the magnetization is very small and 
$\tau$ is about $2/3\approx0.67$. On the other hand,
 at very low-temperatures we have a completely ordered lattice 
with all sites spin up or down, i.e. $|M|\approx1$. 
We emphasize that, for $L=0$, the model reduces to the Blum-Capel model
\cite {RefJ13}
 and corresponding mean-field expressions for 
free energy, magnetization, and $\tau$ can be simply found 
 from (\ref {E:F_mf}),
(\ref {E:M_mf}) and (\ref {E:sigma_mf}). 
In particular, for $H=0$ we get
\begin{eqnarray}\label{E:MBC_mf}
M=\frac {2e^{\beta D}  \sinh{[\beta zJM]}}{1+2e^{\beta D} \cosh{[\beta zJM]}}.
 \end{eqnarray}
Equation (\ref {E:MBC_mf}) enables us to calculate 
mean-field approximation for transition temperature, $T_0$; it's enough to 
expand the expression on the right hand of (\ref {E:MBC_mf}) 
up to the first order of $M$.
As $ M \to 0$ we have $ \sinh [\beta zJM] \to \beta zJM$ and 
$ \cosh [\beta zJM] \to 1 $, thus
\begin{eqnarray}\label{E:T0_mf}
zJ\beta_0=1+\frac {1}{2} e^{-\beta_0 D},
 \end{eqnarray}
where $\beta _0= \frac {1}{k_BT_0} $. Equation (\ref {E:T0_mf})  
determines the curve of phase transition
in phase diagram of Blum-Capel model at $H=0$ plane, 
however up to now we do not know the type of 
phase transition occurring at $ T_0 $. 
Using (\ref {E:M_mf}) we write down the first few terms of 
series expansion of the zero external field free energy around $M=0$:
\begin{eqnarray}\label{E:Fexp_mf}
F_{\rm{MF}} (H=0)\simeq a_0+a_2 M^2+a_4 M^4,
\end{eqnarray}
with
\begin{eqnarray}\label{E:a0_mf}
a_0=-Nk_B T \ln{(1+2e^{\beta D})},                                  
\end{eqnarray}
\begin{eqnarray}\label{E:a2_mf}
a_2=\frac {zJ}{2} [1-\beta zJ \frac {2e^{\beta D}}{1+2e^{\beta D}}],                               
\end{eqnarray}
\begin{eqnarray}\label{E:a4_mf}
a_4=\frac {zJ}{24} {(\beta zJ)}^{3}  \frac {2e^{\beta D}}{1+2e^{\beta D}} [\frac {6e^{\beta D}}{1+2e^{\beta D} }-1].
\end{eqnarray}
The essential condition for having a critical phase transition 
according to Landau theory is \cite {RefB1}
\begin{eqnarray}\label{E:Lnd_2nd}
a_2=0     ,\    a_4>0.                                               
\end{eqnarray}
Hence, to have a critical phase transition we get 
according to mean-field theory 
\begin{eqnarray}\label{E:D_2nd}
D>-zJ \frac {\ln{4}}{3}.  
\end{eqnarray}
In addition, at the tricritical ($\rm{tc}$) point 
where the three phases predicted by 
the classical spin-1 Ising model become critical simultaneously 
we must have
\begin{eqnarray}\label{Lnd_Tri}
a_2=0     ,\     a_4=0.
\end{eqnarray}
Then, the $\rm{tc}$ point is determined by
\begin{eqnarray}\label{Tri_mf}
\beta_{\rm{tc},\rm{MF}}^{(3)} zJ=3        ,\        D_{\rm{tc},\rm{MF}}^{(3)}=-zJ \frac {\ln{4}}{3}.
\end{eqnarray}
\begin{figure}
\begin{center}
  \includegraphics[width=0.45\textwidth]{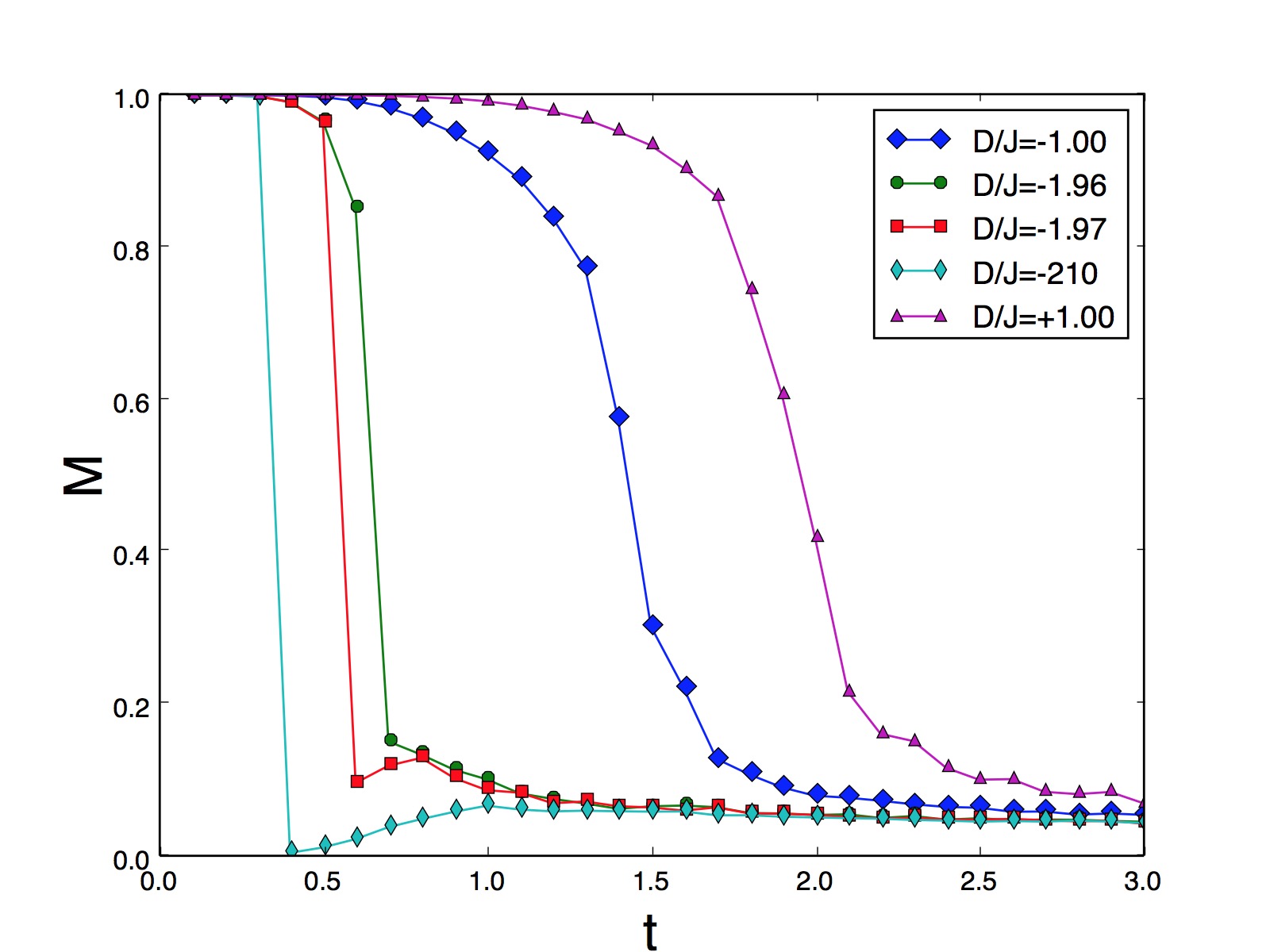}
\caption{Absolute magnetization $M$ vs. reduced temperature 
$t=k_BT/J$ for different values of $D$ for a 2D 
square lattice with $20\times20$ 
sites described by the Blum-Capel Hamiltonian in zero external field 
according to Metropolis Monte Carlo simulation for $D/J=1.00$, $D/J=-1.00$, $D/J=-1.96$, $D/J=-1.97$ and 
 $D/J=-2.10$. The critical phase transition becomes 
first-order at $D_{\rm{tc,MC}}^{(3, \rm{square})}/J=-1.96\pm0.01$, 
 and $t_{\rm{tc,MC}}^{(3, \rm{square})}=0.64\pm0.01$. The error bars are smaller than size of the markers.\label{fig:MCBC3}}        
\end{center}
\end{figure}

Mean-field solution in general is not the exact solution 
but only approximated because it
neglects the effect of dimensionality. 
The results of mean-field calculations become more 
precise when the dimensionality of the system 
becomes larger and the mean-field predictions, like  
for example the critical exponents, become exact if the 
dimensionality of the system is equal or higher than the 
upper critical dimension, $d_{\rm{up}}$, which is given by \cite {RefJ18_new} 
\begin{eqnarray}\label{E:Up-D}
d_{\rm{up}}=\frac{(\gamma+2\beta)}{\nu}.
\end{eqnarray}

Mean-field usually gives good predictions for 
the phase diagrams of the 3D systems \cite {RefB1}. 
In this respect, the interesting example is a 3D system like a cubic lattice 
at the $\rm {tc}$ point. At the $\rm {tc}$ point 
we have the following critical exponents \cite {RefJ18_new}
\begin{eqnarray}\label{E:Crit-t}
\beta_ {\rm {tc}}=\frac{1}{4},\ \nu_ { \rm{tc}}=\frac{1}{2},\ \gamma_ {\rm{tc}}=1,
\end{eqnarray}
(\ref {E:Up-D}) and (\ref {E:Crit-t}) lead to
 \begin{eqnarray}\label{E:Dup-value}
d_{\rm{up}}=3.
\end{eqnarray}
It means that at the $\rm {tc}$ point, mean-field theory 
provides a very good description of the model in
 3D lattices in terms of critical exponents. So for the spin-1 3D Ising model, 
 which is described by the Hamiltonian (\ref {E:Hamil3}), 
around the $\rm{tc}$ point, for zero external field 
 and $L=0$ the critical exponents are the ones given by (\ref {E:Crit-t}).  

\begin{figure}
\begin{center}
  \includegraphics[width=0.47\textwidth]{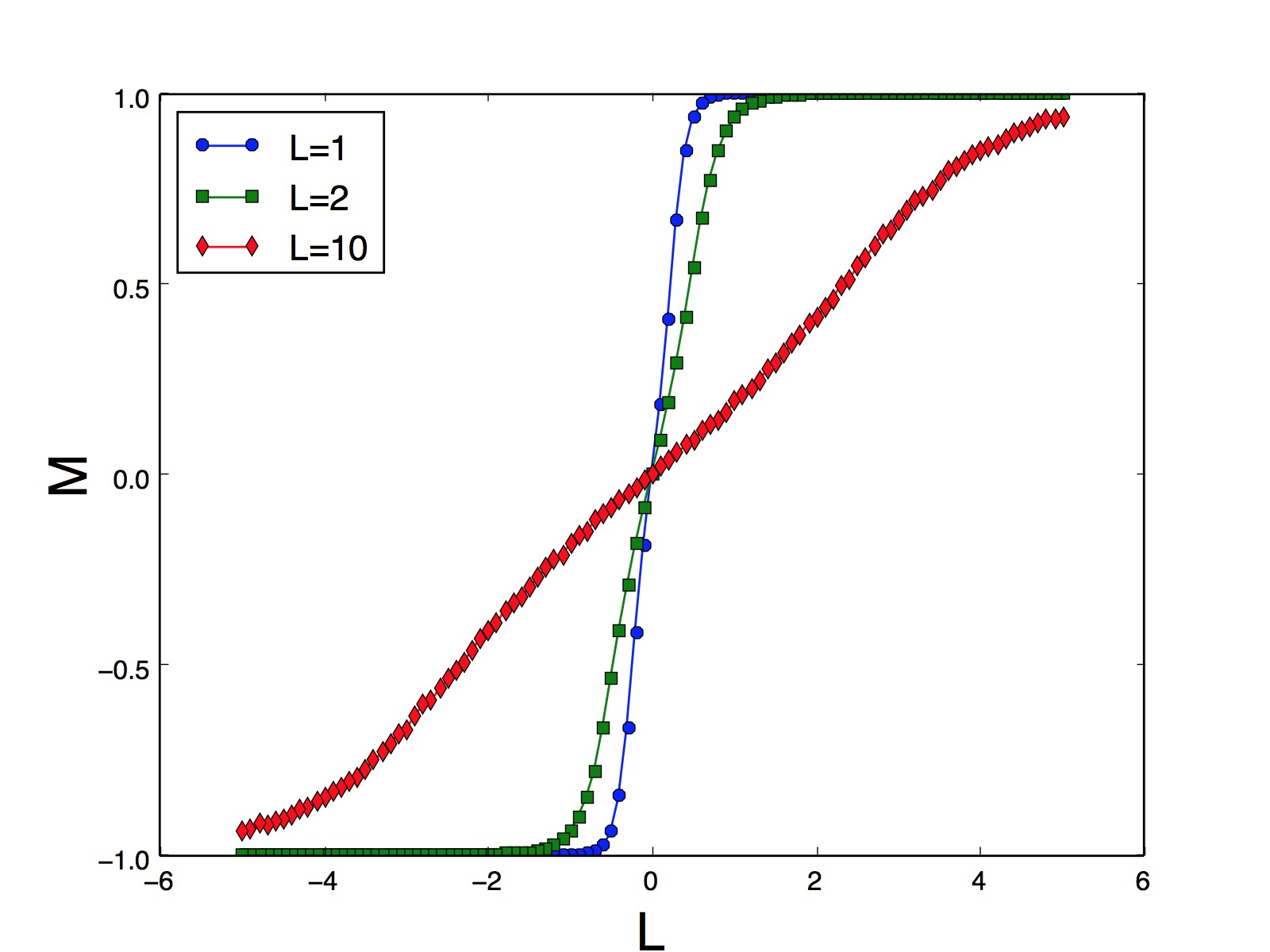}
\caption{Magnetization as a function of  L plots obtained from 
Metropolis Monte Carlo simulation for the spin-1 Ising model 
defined by Hamiltonian (\ref {E:Hamil3}) for a $20\times20$ 
square lattice with $H=D=J=0$ for $k_BT/L=1.0$, $k_BT/L=2.0$, \emph{Bottom}: $k_BT/L=10.0$,
 which indicates that no phase transition takes place. Error bars are smaller than size of the markers.} 
\label{fig:ML-3}       
\end{center}
\end{figure}

Now we use the Metropolis Monte Carlo technique 
to investigate the behavior of 2D square and 3D cubic lattices defined 
by the Hamiltonian (\ref {E:Hamil3}) and compare 
the results with those of mean-field theory. 
Figure \ref {fig:MCBC3} shows the absolute magnetization 
versus reduced temperature 
for different values of $D$ obtained by Metropolis Monte Carlo 
simulation for a 2D square lattice. 
It agrees qualitatively with
mean-field but obviously leads to different values for the $\rm{tc}$ point:
\begin{align}\label{Tri_MC}
\frac {D_{\rm{tc},\rm{MC}}^{(3,\rm{square})}}{J}&=-1.96\pm0.01,\nonumber \\        \frac {k_BT_{\rm{tc},\rm{MC}}^{(3,\rm{square})}}{J}&=0.64\pm0.01.  
\end{align}
We have already seen that mean-field approximation suggests 
that the spin-1 Ising model governed by 
(\ref {E:Hamil3}) does not exhibit any phase transitions when 
$L$ is nonzero.Interestingly, Metropolis Monte Carlo simulation 
that is a more accurate method confirms this result.
For instance, Figure \ref {fig:ML-3}  proves the non-existence of 
the phase transition for a 2D square lattice with $20\times20$ spins
in the case in which only $L$ is non-zero. For a 3D cubic lattice
the system behavior is similar and $\rm{tc}$ point is
given by
\begin{align}\label{Tri_MC-cube}
\frac {D_{\rm{tc},\rm{MC}}^{(3,\rm{cube})}}{J}&=-2.86\pm0.01,\nonumber \\        \frac {T_{\rm{tc},\rm{MC}}^{(3,\rm{cube})}}{J}&=1.4\pm0.1.  \end{align}

\subsection{Case 4}
\label{sec:3.4}
As another specific case we consider the following spin-1 Ising model 
\begin{eqnarray}\label{BEG}
{\mathcal{H}_{\rm{Ising1}}^{(4)}}
=-J\sum_{<ij>}s_i s_j -K\sum_{<ij>}s_i^2 s_j^2 -D\sum_{i=1}^{N}s_i^2. \nonumber \\
\end{eqnarray}     
First note that this Hamiltonian is equivalent to the Ising 
spin-1/2 lattice gas with following Hamiltonian:    
\begin{eqnarray}\label{Lgas}
\mathcal{H}_{\rm{lg}}
=-J\sum_{<ij>}s_i s_j t_i t_j -K\sum_{<ij>}t_i t_j -D\sum_{i=1}^{N}t_i, \nonumber \\ 
\end{eqnarray}     
where $s_i=\pm1$ and $t_i=0,1$ and the subscript $lg$ denotes lattice gas. 
Before we prove this equivalence let us discuss about the Hamiltonian 
(\ref {Lgas}) shortly. For simplicity, we assume that we have a 
2D square lattice with $N$ sites. According 
to the spin-1/2 Ising lattice gas model each site can be occupied 
with a particle or it can be a vacancy. 
If site $i$ is occupied, the variable $t_i$ is one, otherwise it will be zero.  
In the Hamiltonian (\ref {Lgas}) the first 
term proportional to $J$ expresses an exchange interaction between 
two neighbor sites if and only if both are occupied and the amount 
and sign of this interaction depend 
on spin variables of these two occupied sites. The second term 
of $\mathcal{H}_{\rm{lg}}$ proportional to $K$ is the interaction energy
 between a pair of filled neighbors regardless of their spins 
and the last term proportional to $D$  
is a spin independent 
 effect of some external field with occupied sites 
with $D$ playing the role of this field. 
Now we are ready to prove 
 the equivalence of 
$\mathcal{H}_{\rm{Ising1}}^{(4)}$ given by
 (\ref {BEG}) and $\mathcal{H}_{\rm{lg}}$ given by (\ref {Lgas}).
 To do it we start from (\ref {Lgas}) and impose the following transformation
\begin{eqnarray}\label{Trans}
r_i=t_i s_i                                                            
\end{eqnarray}
Equation (\ref {Trans}) illustrates that $r_i$ can be $+1$, $-1$, or $0$. 
Obviously $r_i^2$  only has two 
 possible values: $+1$, or $0$. So in the two last terms of (\ref {Lgas}) 
we can substitute $t_i$ with $r_i^2$. 
 Thus $\mathcal{H}_{\rm{lg}}$ in terms of new spin variable $r_i$ is
\begin{eqnarray}\label{AfterTrans}
\mathcal{H}_{\rm{lg}}
 =-J\sum_{<ij>}r_i r_j -K\sum_{<ij>}r_i^2 r_j^2 -D\sum_{i=1}^{N}r_i^2.                      
\end{eqnarray}
Thus Hamiltonians (\ref {BEG}) and (\ref {Lgas}) are equivalent 
and share the same physics. 
This equivalence is conceptually trivial, because spin-1 
 model can be considered as a spin-1/2 model with vacancies 
but the underlying physics is interesting.
Regarding this point, historically Blume, Emery, and Griffiths
 suggested the Hamiltonian
 (\ref {BEG}) as a spin-1 lattice model to describe a mixture 
 of non-magnetic $(s = 0)$ and magnetic 
$(s = \pm1)$ components \cite {RefJ7}. The model was 
 originally inspired by the experimental observation 
that the continuous superfluid transition in $\rm{He}^3$ with 
 $\rm{He}^4$ impurity becomes a first order transition into normal 
and superfluid phase separation above 
 some critical $\rm{He}^3$ concentration. 
\begin{table}
\caption{Metropolis Monte Carlo simulation result 
for $T_{\rm{tc}}$ and $D_{\rm{tc}}$ for 
different values of $K$ for a $20\times20$ 2D spin square lattice 
governed by the 
Hamiltonian (\ref {BEG}). }
\label{tab:2}       
\begin{center}
\begin{tabular}{|c|c|c|}
\hline
$K/J$ & $k_BT_{\rm{tc}}/J$ & $D_{\rm{tc}}/J$  \\
\hline
0.00 & 0.64 & -1.96  \\
0.10 & 0.68 & -2.16  \\
0.20 & 0.75 & -2.36 \\
0.30 & 0.82 & -2.56 \\
0.40 & 0.85 & -2.75 \\
0.50 & 0.92 & -2.96 \\
0.60 & 0.97 & -3.16 \\

\hline
\end{tabular}\\
\end{center}
\end{table}
\normalsize
Blume, Emery and Griffiths have found the mean-field solution and have 
determined the approximated
phase diagram and the $\rm{tc}$ of the model. 
Since we have found that Metropolis Monte Carlo results qualitatively agree
with the mean-field solution, e.g. phase diagrams 
obtained from the two methods are similar, 
in Table \ref {tab:2} we present the Metropolis Monte Carlo simulation
results for the tc temperature $T_{\rm{tc}}$ and 
the tc crystal field of strength $D_{\rm{tc}}$ for
a 2D square lattice. These results show that with increasing $K$ there is
an increase of  $T_{\rm{tc}}$ and a negative  increase of $D_{\rm{tc}}$.
\subsection{Case 5: long-range Ising spin-1 model}
\label{sec:3.5} 
Long-range interaction that are typical of statistical mechanical systems
may affect the critical behavior 
of the corresponding models \cite{RefJ16}. 
Regarding this, in this section we deal with the 2D spin-1 Ising model 
with long-range spin interactions. We investigate the effect 
of this further interaction using Metropolis Monte 
Carlo simulation comparing the results with the ones of the 
corresponding 2D spin-1/2 Ising model in the presence of the same interaction. 
In analogy with long-range spin-1/2 Ising model \cite {RefJ17,RefJ18} we define 
the long-range Hamiltonian for spin-1 Ising model as follows 
\begin{eqnarray}\label{LRIH}
\mathcal{H}_{\rm{lr}}^{(5)}
=-\sum_{ij}\frac {J}{r_{ij}^{d+\sigma}} s_i s_j,                       
\end{eqnarray}
where $s_i=0$, $1$, or $-1$, $d$ is the lattice dimensionality, 
$\sigma$ is the phenomenological parameter 
which determines the interaction strength, $r_{ij}$ is the distance between a couple of spins labeled by the indices 
$i$ and $j$, and the subscript $lr$ denotes long range. 
Explicitly, in the Metropolis algorithm if the 
flipped spin is at position $(x,y)$ we have   
\begin{eqnarray}\label{rij}
r_{ij}=\sqrt{{(x-i)}^2+{(y-j)}^2}.      
\end{eqnarray}  
In this analysis we limit ourselves to ferromagnetic 
materials, i.e. $J>0$.
 Notice that for spin-1/2 Ising model the Hamiltonian  (\ref {LRIH}) 
has the same expression but  $s_i=1$, or $-1$.  
\begin{figure}
\begin{center}
  \includegraphics[width=0.45\textwidth]{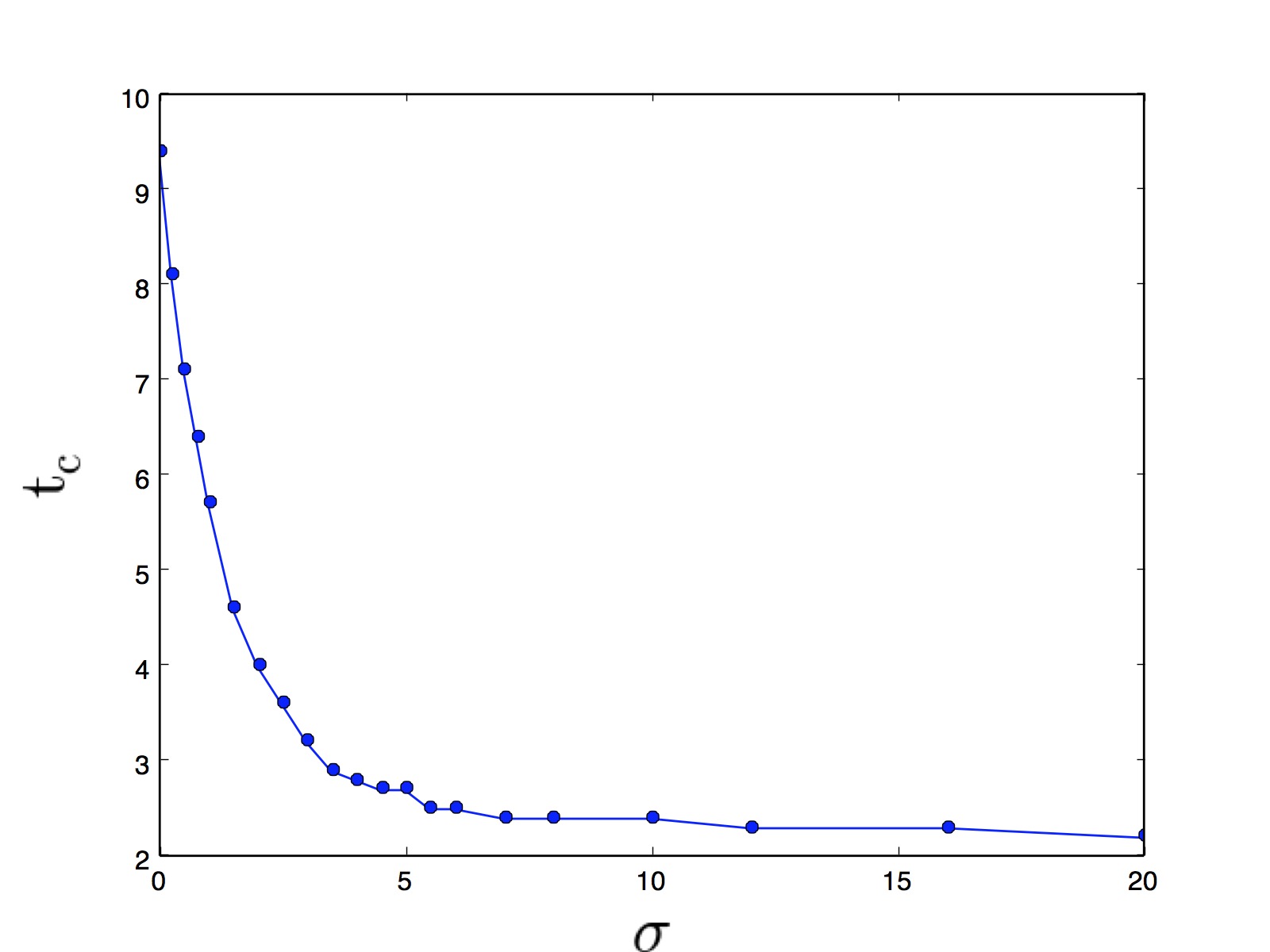}
  \includegraphics[width=0.45\textwidth]{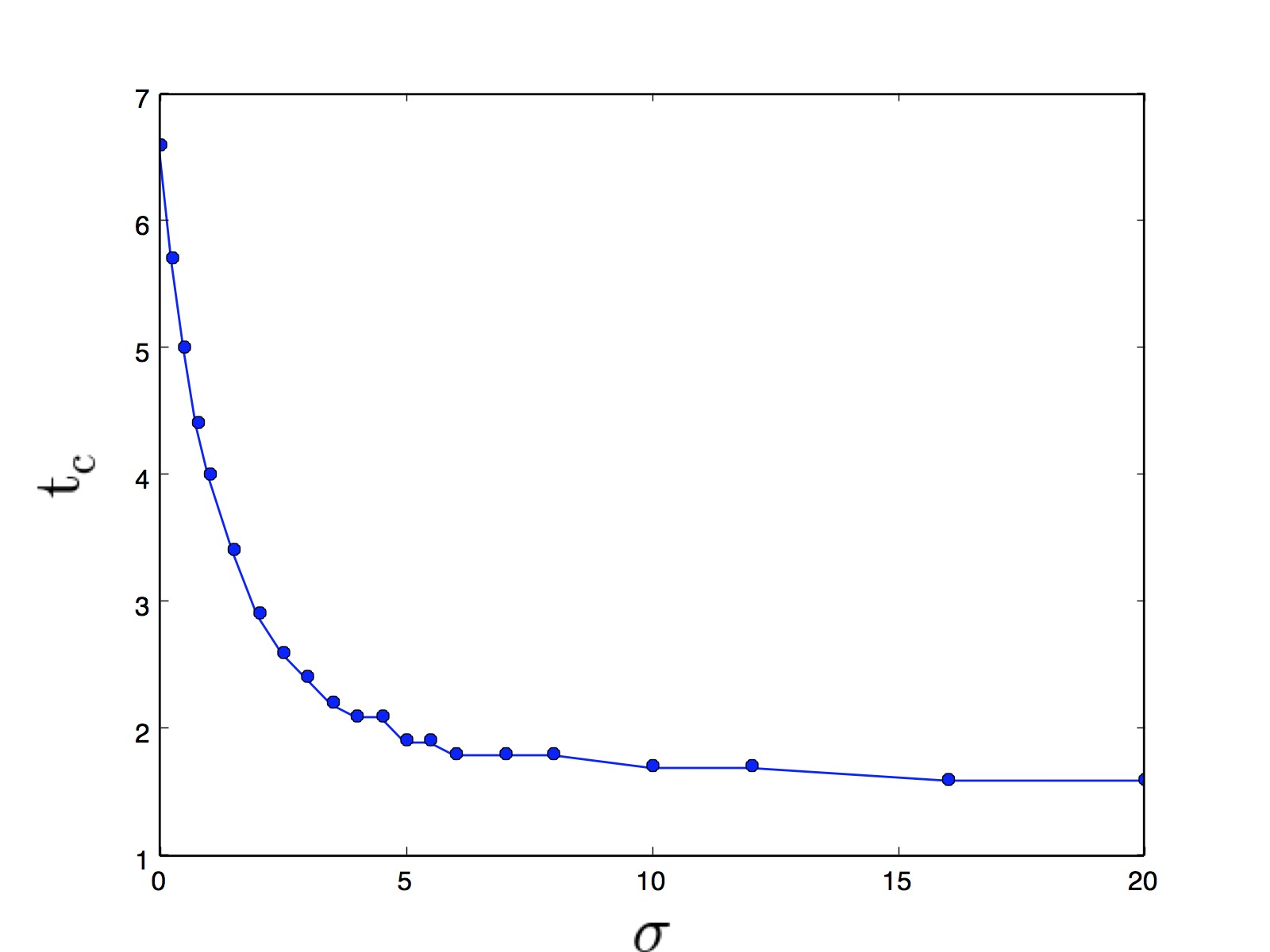}
\caption{$t_c=\frac{k_B T_c}{J}$ as a function of $\sigma$ for 
the long-range interaction Ising model \emph{Top}: spin-1/2, \emph{Bottom}: spin-1 for a square lattice with $40\times40$ sites 
described by Hamiltonian (\ref {LRIH}) with $R=7$ 
obtained from Metropolis Monte Carlo simulation. Error bars are smaller than size of the markers.} 
\label{fig:Tc-4}       
\end{center}
\end{figure}
On the basis of this numerical simulation we can investigate 
the dependence of the critical temperature on $\sigma$.
We do the Metropolis Monte Carlo simulation 
for different values of parameter $\sigma$, considering 
the long-range interaction and assuming 
that each spin interacts with other spins which their 
distance is equal or less than some radius $R$. 
In other words the summation in the Hamiltonian 
(\ref {LRIH}) is carried out over all spins, 
which are in the circle of radius $R$ around the flipped spin 
in each step of Metropolis algorithm. It means that in the algorithm $r_{ij}<R$.
Figure \ref{fig:Tc-4} shows the result for $R=7$. 
As we expect the critical temperature of the model decreases 
with parameter $\sigma$. The dependence of 
critical temperature on parameter $\sigma$, basing upon 
the results of numerical modelling of the least squares for spin-1/2 is given by
\begin{eqnarray}\label{t_c-sigma-half}
\frac{k_B T_{c,\rm{spin-1/2}}^{\rm{lr}}}{J}=2.3+6.9e^{-0.7\sigma}.                                            
\end{eqnarray}
Similarly for spin-1 model we get:
\begin{eqnarray}\label{t_c-sigma-one}
\frac{k_B T_{c,\rm{spin-1}}^{\rm{lr}}}{J}=1.7+4.7e^{-0.7\sigma}.                                         
\end{eqnarray}
So in both cases we have 
\begin{eqnarray}\label{t_c-sigma-half_one}
T_c=a+be^{-c\sigma},                                             
\end{eqnarray}
where $a$ is the critical temperature of the short-range 
model and $c\approx0.7$.
\section{Conclusion}
\label{sec:4}

In the present work we studied the classical spin-1 Ising model  
using different analytical and numerical methods 
such as mean-field theory, series expansions and 
Monte Carlo simulation to investigate some
critical properties of the model like critical temperature 
and critical exponents for 1D chain,
2D square lattice, and 3D cubic lattice. 
We have found that, albeit some similarities with 
the critical behavior of the classical spin-1/2 Ising model, 
because of the presence in the
Hamiltonian of the spin-1 model of more terms, i.e. more different 
types of interactions between spin pairs, the critical properties of this model are much richer and more variegated with respect to 
the ones of the corresponding spin-1/2 Ising model.

We have used mean-field theory that represents a 
strong mathematical tool to study 
the physics of the model in some special cases. We 
have found that,
for 3D lattices near the tricritical point, the critical properties
of the model can be described 
by mean-field theory. In particular, we have found that the 
critical exponents around the tricritical point 
calculated via the mean-field approximation 
are confirmed by Monte Carlo simulations. 
On the other hand, the mean-field results for 2D lattices are 
only qualitatively but not quantitatively
correct as highlighted by Monte Carlo simulation. 
The simulation results obtained
for 2D square and 3D cubic lattices can be easily 
extended to other types of lattices.

We have shown that, for a special case of 
the spin-1 Ising model 
where the bilinear and the bicubic terms 
are set equal to zero,
it is possible to write the corresponding
partition function in arbitrary 
dimensions as the one of the spin-1/2 Ising 
model in agreement with our Monte Carlo simulation.
Finally, we have investigated the long-range 
spin-1 Ising model Hamiltonian by 
including in the Hamiltonian a long-range interaction 
term in analogy with what was carried out for spin-1/2 Ising 
model determining the dependence of the critical 
temperature of 
the two models on the strength of this interaction. 

 \section*{Acknowledgements}   
This work was partially supported by National Group 
of Mathematical Physics (GNFM-INdAM) 
and Istituto Nazionale di Alta Matematica “F. Severi”. 

\bibliographystyle{apsrev4-1}
\bibliography{refs.bib}

\end{document}